# LATE—TIME ANISOTROPY AND RADIATION DRAG WITHIN THE COSMIC MICROWAVE BACKGROUND

M. Kuassivi
170374 Cotonou, Benin

ABSTRACT

I investigate the effect of the radiation drag force on a star moving relative to the Cosmic Microwave Background (CMB) at z = 0. As expected, the perturbation is extremely small and has no observable consequences on orbital motions of stars within a galaxy or on orbital motions of galaxies within a cluster. The energy dissipated by cubic meter in a galaxy via this mechanism is $10^{14}$ smaller than the energy density of the CMB and, thus, is a negligible source of anisotropy. However, from the last scattering surface to population III stars (30 < z < 1000), radiation drag on particles may have played a major role in the apparition of anisotropy and rapid formation of densities.

Keywords :

1. INTRODUCTION

The Solar System has been a domain of important discoveries, which have helped us to justify and verify basics laws of physics. The kinematics of planetary motion, described by Johannes Kepler (1571—1630) at the beginning of the $17^{th}$ century , and the corresponding dynamics, unveiled by Isaac Newton (1642—1727) at the end of the $17^{th}$ century, are the best known examples.

For a long time, the Newtonian gravitational attraction has been thought to be the only driving force affecting the motion of celestial bodies. On the other hand, Kepler already speculated about some kind of repulsive force, because he knew the tail of comets are directed away from the Sun.

The precise observations of Solar System objects in the $20^{th}$ century allowed us to not only verify the current theory of the gravitation, the General Theory of Relativity, but also to recognise the role of electromagnetic forces, which subtantially affect the dynamics of small objects.

We can distinguish a number of modes how the electromagnetic field interacts with matter: cometary jets, radiation pressure, Poynting—Robertson effect, the Yarkovsky/YORP effect, and the Lorentz force.

Tiny non—gravitational effects may become important in the dynamics of large bodies provided that the effects leads to long—term perturbation accumulating over long time span (Vokrouhliky & Milani, 2000).

Radiation force effect in the motion of the moon represent an outstanding example: absorbed, reflected and thermally—reprocessed sunlight cause a synodic oscillation of the lunar geocentric distance of about 4 millimeters (Vokrouhliky, 1997).

The goal of this paper is to investigate the so far neglected effect of CMB radiation pressure on large systems (stars, galaxies).

The paper is organized as follows: in Sect. 2, I derive the drag force acting on a large spherical mirror moving through a thermal radiation bath. In Sect. 3, I show the expected order of magnitude and character of the perturbation in the case of a stellar object, a galaxie and a cluster of galaxies. Sect. 4 contains a short discussion on the possible consequences of this mechanism in the past.

## 2. THEORY

A black or mirror sphere of surface area A moving at velocity v in a bath of thermal radiation at temperature T will eventually stop (W.D. Smith 1995). The idea is that the sphere sees a blue shifted (to the front, or redshifted, to the rear) blackbody radiation and hence experiences a net force. This can easily be generalized to stars moving through the CMB assuming that: (1) the stellar atmosphere reflects the CMB (essentially made of radio—waves), (2) the stellar radius is much larger than the thermal wavelength, (3) the star is not set into vibration by the incident waves, and (4) the time—scales are much larger than the gaps between collisions with single photon.

Let us consider a plane mirror moving with velocity v through a thermalized photonic gas with temperature T. In the frame of the miror, the distribution function of photons is given by the expression (Lightman, Press, Price & Teukolsky 1975):

$$(1) \quad n\left(\vec{k}, \vec{v}\right) = \frac{1}{e^{\left(\gamma \hbar \left(\omega + \vec{k} \cdot \vec{v}\right)/kT\right)} - 1}$$

where

$$(2) \quad \gamma = 1/\sqrt{1 - \beta^2}, \quad \beta = v/c$$

By using the distribution function we find the momentum density P of the elecromagnetic field (Balasanyan & Mkrtchian, 2009)

$$(3) \vec{P} = -2\hbar \iiint \vec{k} \cdot n\left(\vec{k}, \vec{v}\right) \frac{dk_x dk_y dk_z}{\hbar^3}$$

$$(4) \vec{P} = -\vec{v} \cdot \frac{16}{3c^3} \sigma T^4 \frac{1}{1 - \beta^2}$$

Thus, for a plane mirror moving perpendicularly to its surface we get the drag force f per unite area of surface

$$(5) f = \frac{32}{3c} \sigma T^4 \frac{\beta}{1 - \beta^2}$$

As seen, the drag force is proportional to $T^4$ as in the case of non moving metallic particle.

From here, the case of the spherical mirror can be derived by integration on the half sphere:

$$(6) F = \iint f \cdot R^2 \cdot \sin^2 \theta \cdot \cos \theta \, d\theta \, d\phi$$

$$(7) F = \frac{128}{9c} \sigma T^4 \frac{\beta}{1 - \beta^2} \pi R^2$$

This is the main result of this communication and it is correct for arbitrary large velocities of the sphere.

## 3. ORBITAL PERTURBATION AT Z=0

Let first consider a solar—type star (R=500000 km) moving at 200 km.s$^{-1}$ through a 3K CMB at 10 000 pc from the galactic center. A rapid computation shows that the net drag force F is extremely weak (of the order of 100 N).
Along one orbit (200 My), the energy dissipated by the star via this mechanism is given by

$$(8) W_{\text{diss}} = F \cdot L_{\text{orb}}$$

which gives Wdiss of the order of a few time $10^{23}$ Joules. Thus during its life time, the Sun will dissipate as much as $10^{25}$ Joules in the CMB.

As a matter of fact, this energy is $10^{17}$ times smaller than the total kinetic energy of the Sun and shows that the mechanism has almost no impact on the energy balance of the orbital motion.

In the case of a large galaxy moving through the intergalactic medium all the star—CMB interactions add up. For a large Galaxy containing $10^{11}$ solar masses travelling 1 Mpc at 200 km.s$^{-1}$, the mean total energy dissipated in the CMB is $E_{diss} \sim 10^{35}$ J. But this value is negligible in comparizon to the external gravitational potential. For instance, the interaction energy between 2 massive galaxies ($10^{11}$ solar masses) distant from 1 Mpc is of the order of $10^{49}$ Joules.

The same conclusion is reached when comparing the gravitational interaction between clusters of galaxies and CMB drag dissipation. A giant cluster of $10^5$ galaxies (each one containing $10^{11}$ solar masses) travelling 300 Mpc at 2000 km.s$^{-1}$ will dissipate as much as $10^{44}$ J in the CMB. But again, this is negligible in comparizon to the gravitational interaction between 2 giant clusters distant from 300 Mpc ($E_{pot} \sim 10^{57}$ J)

## 4. DISCUSSION

In the previous section, I have shown that the CMB radiation drag has no measurable influence on stars or galaxies because the temperature of it is too low. On the other hand, we may ask wether the energy balance of the CMB is impacted by the dissipation. In the following, I consider a massive ($10^{11}$ solar masses) comoving galaxy assuming that it can be approximated by a thick (1 kpc) disk with a radius of 20 kpc and I compute the total energy dissipated locally in the CMB during 10 billions years by the stars. The total energy dissipated is of the order of $10^{36}$ J. A rapid computation of the total CMB energy within that same volume (I assume an energy density of 4 $10^{-14}$ J/m$^3$) gives $10^{48}$ J. We can thus conclude that the effects of radiation drag (a few times $10^{-14}$ K) on the CMB cannot yet be observed since WMAP has a sensitivity of a few μK (Jarosik et al. 2007).

Clearly, CMB photons do not shape the current Universe. But they might have shaped it moments after the decoupling (z ~ 1000) through an enhanced dissipation mechanism. At that time the CMB temperature reached 1000 K in which case all our estimates would all be multiplied by $10^{12}$. As a result, intense dissipation of kinetic energy and accelerated loss of angular momentum could have led to the formation of stars and galaxies. Radiation drag (Poynting—Roberston effect) has already been invoked for the formation of supermassive black holes in disk galaxies by Kawakatu & Umemura (2004). If such is the case, we might also expect to see traces of this past interaction in the current background anisotropy.